# The Strength and Radial Profile of Coronal Magnetic Field from the Standoff Distance of a CME-driven Shock

Nat Gopalswamy[1] and Seiji Yashiro[2]

[1]NASA Goddard Space Flight Center, Greenbelt, MD 20771, USA

[2]The Catholic University of America, Washington DC 20064, USA

**ABSTRACT**

We determine the coronal magnetic field strength in the heliocentric distance range 6 to 23 solar radii (Rs) by measuring the shock standoff distance and the radius of curvature of the flux rope during the 2008 March 25 coronal mass ejection (CME) imaged by white-light coronagraphs. Assuming the adiabatic index, we determine the Alfven Mach number, and hence the Alfven speed in the ambient medium using the measured shock speed. By measuring the upstream plasma density using polarization brightness images, we finally get the magnetic field strength upstream of the shock. The estimated magnetic field decreases from ~48 mG around 6 Rs to 8 mG at 23 Rs. The radial profile of the magnetic field can be described by a power law in agreement with other estimates at similar heliocentric distances.

*Subject headings*: sun: coronal mass ejections — sun: magnetic field — sun: corona

## 1. **Introduction**

Magnetic field strength in the solar atmosphere is routinely measured at present only in the photospheric and chromospheric layers. The coronal magnetic field is estimated from the photospheric and chromospheric values using extrapolation techniques (see e. g., Wiegelmann, 2008 and references therein). Direct measurement of coronal magnetic fields is possible at microwave (see e.g., Lee, 2007) and infrared (Lin et al., 2004) wavelengths, but these correspond to regions very close to the base of the corona. The extrapolation methods involve assumptions such as low-beta plasma, which may not be valid in the outer corona (Gary 2001). Faraday rotation techniques have also been used in estimating the magnetic field strengths at several solar radii from the Sun center (Pätzold et al., 1987; Spangler, 2005; Ingleby et al., 2007). In this paper, we describe a new technique to measure the coronal magnetic field that makes use of the white-light shock structure of coronal mass ejections (CMEs) observed in coronagraphic images (Sheeley et al., 2000; Vourlidas et al., 2003; Gopalswamy, 2009; Gopalswamy et al., 2009; Ontiveros and Vourlidas, 2009). The technique involves measuring the shock standoff distance and the radius of curvature of the driving CME flux rope, which are related to the upstream shock Mach number. Once the Mach number is known, the Alfven speed can be derived using the measured shock speed and hence the magnetic field using a coronal density estimate. The shock can be tracked for large distances within the coronagraphic field of view and hence we obtain the radial profile of the coronal magnetic field. Previous works involving white-light shock structure (Bemporad & Mancuso, 2010; Ontiveros & Vourlidas, 2009; Eselevich & Eselevich, 2011) mainly used the density compression ratio across the shock to derive the shock properties. To our knowledge, this is the first time the shock standoff distance is used to measure the magnetic field in the outer corona.

## 2. Observations

In a recent paper, Gopalswamy et al. (2009) reported on the 2008 March 25 CME, which clearly showed all the CME substructures: the shock sheath, CME flux rope, and the prominence core. The CME was observed by the Sun Earth Connection Coronal and Heliospheric Investigation (SECCHI, Howard et al., 2008) coronagraphs on board the Solar Terrestrial Relations (STEREO, Kaiser et al., 2008) mission. The early phase of the shock surrounding the CME was observed by the Extreme Ultraviolet Imager (EUVI) on board STEREO. The CME was also imaged by the Large Angle and Spectrometric Coronagraph (LASCO, Brueckner et al., 1995) telescopes C2 and C3 on board the Solar and Heliospheric Observatory (SOHO) mission. The STEREO-ahead (SA) spacecraft was ~ 24° ahead of Earth while STEREO-behind (SB) was ~ 24° behind Earth at the time of the eruption. Thus, the east-limb eruption (S13E78) in Earth view corresponds to ~E102 and E54 in SA and SB views, respectively. Therefore, measurements made in the sky plane from SA and Earth views have minimal projection effects. We combine these measurements for the purposes of this paper.

The type II radio burst observed during the eruption indicates the formation of the shock when the CME was at a heliocentric distance of ~1.5 Rs. However, the type II bursts ended when the CME was at ~3.7 Rs, beyond which the shock existed but was radio quiet. This means the shock must have attained the subcritical regime. The CME first appeared in the LASCO/C2 field of view at 19:31 UT, when the shock was already at a heliocentric distance of 5.9 Rs. However, LASCO/C3 tracked the CME flux rope until it reached a distance of ~23Rs. SECCHI/COR2A observed the flux rope and shock in the intermediate distance range: 2.3 to 11.51 Rs, but the shock and flux rope structures are clearly visible only from 6.5 Rs onwards. The SECCHI/COR1A also observed the shock, but the shock structure is seen only at the flanks, so we do not use this data. In all, we have shock – flux rope measurements at 10 different heliocentric distances from ~6 Rs to 23 Rs, over a period of ~3 h. These measurements are adequate to obtain the strength and radial profile of the magnetic field over a distance range that exceeded previous ranges (Dulk and McLean, 1978; Pätzold et al., 1987). Figure 1 shows the diffuse shock sheath that surrounds the flux rope at two instances in the STEREO/COR2 and SOHO/LASCO images. The thickness of the shock sheath is the standoff distance. The circle fit to the CME flux rope is also shown.

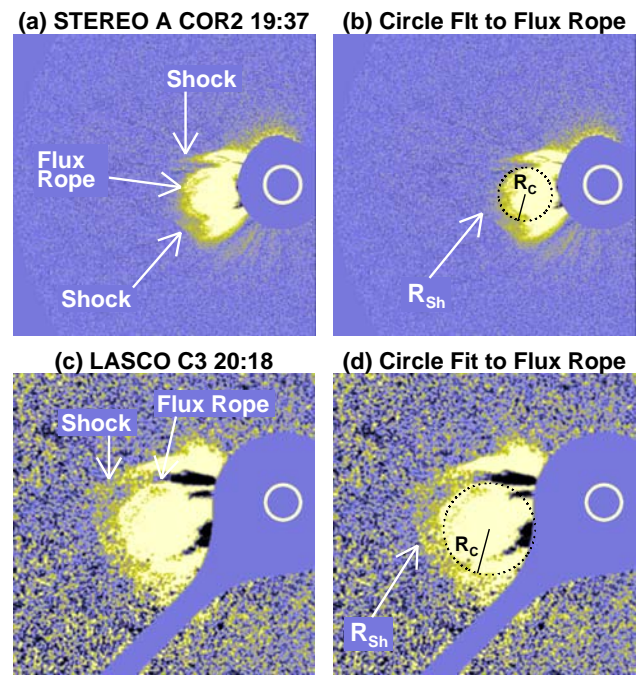

Figure 1. A STEREO-A COR2 (a) SOHO/LASCO/C3 (c) difference images at 19:37 and 20:18 UT showing the shock and the flux rope with the circles fit to the flux rope superposed in (b,d). Images at 19:23 UT (COR2) and 19:43 UT (C3) were used for differencing. The occulting disk blocks the photosphere (represented by the white circle); the pylon extends to the southeast in (c,d). The flux rope radius $R_c$ increases from 1.5 Rs at 19:37 UT to 2.65 Rs at 20:18 UT.

### 3. Analysis and Results

Russell and Mulligan (2002) derived the following relation between the standoff distance $\Delta R$ of an interplanetary shock and the radius of curvature ($R_c$) of the driving CMEs at 1 AU:

$$\Delta R/R_c = 0.81[(\gamma-1)M^2 + 2]/[(\gamma+1)(M^2-1)], \ldots\ldots\ldots\ldots\ldots\ldots\ldots\ldots\ldots\ldots\ldots(1)$$

where M is the shock Mach number and $\gamma$ is the adiabatic index. We apply eq. (1) to CMEs in coronagraphic images because $\Delta R$ is the difference between the shock ($R_{sh}$) and the flux rope ($R_{fl}$) heights from the Sun center. $R_c$ is obtained by fitting a circle to the flux rope (see Fig. 1b,d). For the CME in Fig. 1(d), $R_{sh} = 10.72$ Rs; $R_{fl} = 9.40$ Rs, $R_c = 2.65$ Rs, so $\Delta R/R_c = 0.50$, which gives $M = 1.76$ for $\gamma = 4/3$ and $M = 1.93$ for $\gamma = 5/3$ in eq. (1). The Alfven speed $V_A = (V_{Sh} - V_{SW})/M$, where $V_{sh}$ and $V_{SW}$ are shock and solar wind speeds. $V_{sh}$ can be obtained from the increase in $R_{sh}$ with time; $V_{SW}$ can be obtained from the speed profile derived by Sheeley et al., (1997):

$$V^2_{SW} (r) = 1.75 \times 10^5 [1 - \exp(-(r-4.5)/15.2)] \ \ldots\ldots\ldots\ldots\ldots\ldots\ldots\ldots\ldots\ldots\ldots(2)$$

A linear fit to $R_{sh}$ – time measurements, gives a constant speed of 1195 km/s. A quadratic fit shows that the shock was decelerating with a local speed of 1201 km/s at $R_{sh} = 10.72$ Rs, which we use for illustration. Equation (2) gives $V_{SW} = 243$ km/s at 10.72 Rs. Thus, $V_A = 544$ km/s for $\gamma = 4/3$ and 497 km/s for $\gamma = 5/3$. Finally, we can get the upstream magnetic field B from

$$V_A = 2.18 \times 10^6 n^{-1/2} B \ \ldots\ldots\ldots\ldots\ldots\ldots\ldots\ldots\ldots\ldots\ldots\ldots\ldots\ldots\ldots\ldots\ldots\ldots\ldots\ldots\ldots(3)$$

where n is the upstream plasma density in $cm^{-3}$ and B is in G.

In order to get the coronal density, we inverted the nearest polarization brightness (pB) image before the eruption available online at: http://lasco-www.nrl.navy.mil/content/retrieve/polarize/

using the Solar Software routine "pb_inverter" (Thernisien and Howard, 2006; Cho et al., 2007). The first LASCO/C3 pB images had artifacts on for 2008 March 24 and 25. The second image on March 25 was not useful because it contained the CME, that too close to the edge of the LASCO field of view. So we used the image at 22:50 UT on March 24. The LASCO/C2 pB image was taken at 15:00 UT on March 25, which had glitches at several position angles of interest and was useful only for a few position angles.

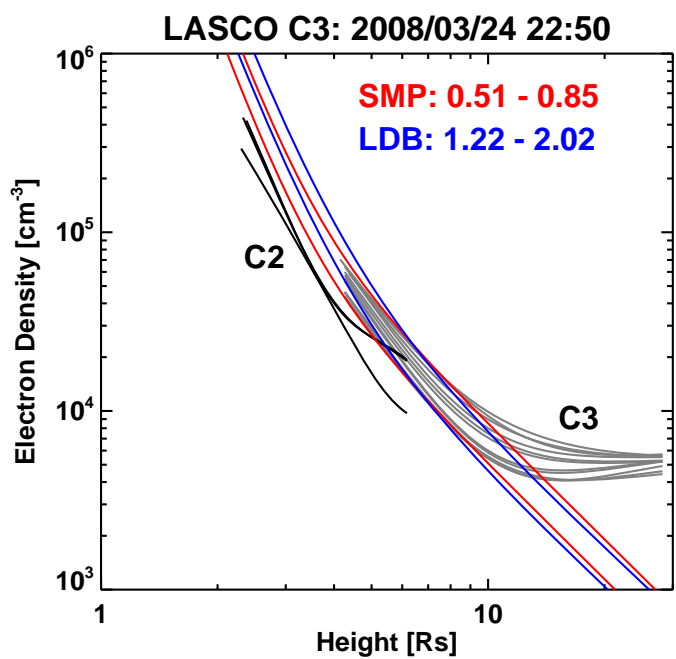

Figure 2. Radial profiles of the electron density at 10 position angles (93° to 103°) around the shock nose from LASCO C3 (gray lines) and C2 (dark lines). Saito, Munro & Poland (SMP) model matches the LASCO C3 density profiles for multiplier of 0.51 to 0.85 (central value 0.68). Leblanc, Dulk & Bougeret (LDB) model matches the LASCO/C3 profiles when multiplied by 1.22 to 2.02 (central value ~1.62).

We selected 10 position angles (93° to 103°) around the shock nose and plotted the density as a function of the heliocentric distance in Fig. 2. The maximum and minimum values give the density range around the shock nose, with the mid value taken as the density at the nose. The C3 pB images yield consistent density values in the range 4 – 9 Rs. Beyond 9 Rs, the pb_inverter program gives a constant density, which is unphysical (see Fig. 2). To get the densities outside the 4 – 9 Rs range, we adjust the Saito, Munro & Poland (1977) (SMP) model,

$$n\,(r) = 1.36 \times 10^6\, r^{-2.14} + 1.68 \times 10^8\, r^{-6.13} \qquad \text{..............................................(4)}$$

and the Leblanc, Dulk & Bougeret (1998) (LDB) model,

$$n\,(r) = 3.3 \times 10^5\, r^{-2} + 4.1 \times 10^6\, r^{-4} + 8.0 \times 10^7\, r^{-6} \quad \text{..............................................(5)}$$

such that the models match the LASCO/C3 densities for certain multipliers. The multipliers corresponding to the central value of the density in the $10^o$ wedge of the C3 pB image at the shock nose are 1.6 for the LDB model and 0.7 for the SMP model. The few position angles that yielded realistic densities from the C2 pB image are consistent with the C3 data (see Fig. 2). For r = 10.72 Rs, density is in the range (4.37 – 7.29) x$10^3$ cm$^{-3}$ with a mid value of 5.83x$10^3$ cm$^{-3}$. With n = 5.83x$10^3$ cm$^{-3}$ in eq. (3) we get B = 19.0±0.53 milligauss (mG) for $\gamma = 4/3$ and 17.4±0.67 mG for $\gamma = 5/3$. The error bars were derived from a combination of the errors in the height measurements and the errors in fitting a circle to the flux rope. Repeating the computation for constant $V_{sh} = 1195$ km/s, we get virtually the same B values: 18.9±0.52 mG for $\gamma = 4/3$ and 17.3±0.67 mG for $\gamma = 5/3$. Linear and quadratic fits to the height-time plot of the shock yield B values that differ by less than 10%.

Following the method outlined above, we computed M, $V_A$, and B at various heliocentric distances in which the shock structure and flux rope were discernible. Table 1 lists the derived and observed quantities along with the uncertainties: UT, observing instrument (SOHO/LASCO or STEREO/COR2), $R_{sh}$, $R_{fl}$, $\Delta R$, $R_c$, $\Delta R/R_c$, M, $V_{sh}$, Vsw, $V_A$, density n from Fig. 2, and finally the magnetic field strength B. The derived values listed in Table 1 are for $\gamma = 4/3$ and the SMP model for extrapolation to larger distances. We also repeated the calculations for $\gamma = 5/3$ and also for the LDB density model. The derived Alfven Mach number is ~ 2 or less implying that the shock was weak as suggested by Gopalswamy et al. (2009). The derived $V_A$ declines from ~660 km/s near 6 Rs to 490 km/s near 23 Rs. Accordingly, the magnetic field declines by an order of magnitude in the heliocentric distance range considered (45.8 ±0.97 mG to 7.58±0.38 mG).

Table 1. Properties of the shock, flux rope and the ambient medium at various heliocentric distances for the 2008 March 25 event assuming $\gamma$=4/3 and SMP density extrapolation

| Time UT | Inst.[a] | $R_{sh}$ $Rs^b$ | $R_{fl}$ $Rs^b$ | $\Delta R$ Rs | $R_c$ $Rs^c$ | $\Delta R/R_c$ | M | $V_{sh}$ km/s | $V_{sw}$ km/s | $V_A$ km/s | N cm$^{-3}$ | B mG |
|---|---|---|---|---|---|---|---|---|---|---|---|---|
| 19:31 | C2 | 5.93±0.14 | 5.08±0.01 | 0.66 | 1.42±0.07 | 0.60 | 1.63 | 1210 | 125 | 664 | 2.26e+04 | 45.8±0.97 |
| 19:37 | CR2 | 6.55±0.05 | 5.86±0.03 | 0.67 | 1.50±0.09 | 0.46 | 1.83 | 1209 | 149 | 580 | 1.77e+04 | 35.4±1.01 |
| 19:42 | C3 | 6.73±0.08 | 6.03±0.05 | 0.75 | 1.71±0.23 | 0.41 | 1.93 | 1208 | 155 | 544 | 1.66e+04 | 32.2±2.32 |
| 20:07 | CR2 | 9.67±0.07 | 8.46±0.06 | 0.95 | 2.39±0.12 | 0.51 | 1.75 | 1203 | 225 | 559 | 7.30e+03 | 21.9±0.49 |
| 20:18 | C3 | 10.72±0.13 | 9.40±0.09 | 1.57 | 2.65±0.16 | 0.50 | 1.76 | 1201 | 243 | 544 | 5.83e+03 | 19.0±0.53 |
| 20:37 | CR2 | 12.50±0.06 | 11.26±0.06 | 1.29 | 3.00±0.25 | 0.41 | 1.92 | 1197 | 268 | 483 | 4.18e+03 | 14.3±0.62 |
| 20:42 | C3 | 13.43±0.19 | 11.40±0.13 | 2.01 | 3.38±0.20 | 0.60 | 1.63 | 1196 | 279 | 562 | 3.58e+03 | 15.4±0.37 |
| 21:18 | C3 | 16.71±0.21 | 14.68±0.19 | 2.25 | 4.00±0.27 | 0.51 | 1.75 | 1190 | 311 | 503 | 2.24e+03 | 10.9±0.34 |
| 21:42 | C3 | 19.54±0.51 | 16.70±0.35 | 2.58 | 4.75±0.38 | 0.60 | 1.63 | 1185 | 332 | 522 | 1.60e+03 | 9.58±0.33 |
| 22:18 | C3 | 22.98±0.39 | 19.84±0.42 | 2.93 | 5.65±0.65 | 0.56 | 1.68 | 1178 | 351 | 492 | 1.13e+03 | 7.58±0.38 |

[a]C2 = LASCO/C2; C3 = LASCO/C3; CR2 = STEREO A/COR2.

[b]Errors in $R_{sh}$ and $R_{fl}$ are the standard deviations of five independent measurements.

[c]Errors in $R_c$ are derived from the circle fitting.

Figure 3 shows the B variation can be fit to a power law of the form,

$$B (r) = pr^{-q}. \quad \dots\dots\dots\dots\dots\dots\dots\dots\dots\dots\dots\dots\dots\dots\dots\dots\dots\dots(6)$$

Using the adjusted SMP model for heights >9 Rs and $\gamma$ = 4/3, we get p = 0.377 and q = 1.25 (data shown in Table 1). The error bars are from height – time measurements and the density range for each height in Fig. 2. For a given density model, the curve (6) becomes slightly flatter for larger $\gamma$ (p = 0.329 and q =1.23). The SMP model extrapolation results in a slightly flatter B profile compared to that from the LDB model, but the difference is almost unnoticeable because the models are normalized to the measured densities in the 4 – 9 Rs range. Note that the STEREO and SOHO measurements yield consistent result.

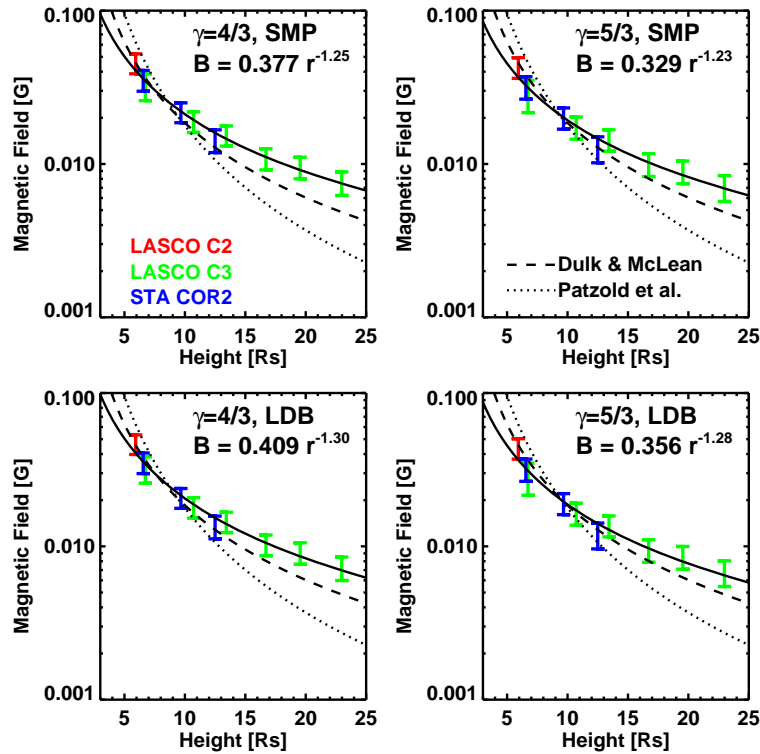

Figure 3. Radial profiles of the magnetic field for γ = 4/3 (left panels), 5/3 (right panels) and two density models (SMP – top panels, LDB – bottom panels). Dulk and McLean (1978) and Pätzold et al. (1987) empirical profiles are shown for comparison. The error bars are a combination of the density range for each height shown in Fig. 2 and the height-time measurement errors. LASCO C2, C3 and STEREO – A COR2 measurements are distinguished using different colors.

Now we compare the magnetic field strengths derived from our technique with those from empirical models and isolated measurements at certain heights. The dashed curves in Fig. 3 is the Dulk and McLean (1978) empirical relation for B above active regions (for r ≤ 10 Rs):

$$B(r) = 0.5 \ (r-1)^{-1.5}. \quad \text{............................................................(7)}$$

From Faraday rotation measurements, Pätzold et al. (1987) derived the profile ($2 \leq r \leq 15$ Rs),

$$B(r) = 6r^{-3} + 1.18r^{-2}, \dots\dots\dots\dots\dots\dots\dots\dots\dots\dots\dots\dots\dots\dots\dots\dots\dots\dots\dots\dots\dots\dots\dots\dots\dots(8)$$

shown as the dotted curves in Fig. 3. We see that the B profile derived from our technique [eq. (6)] is flatter than both these empirical profiles, difference between being larger at larger heights. For example, at r = 23 Rs, eq.(6) gives 6.9 mG for $\gamma = 4/3$ and LDB extrapolation; the Dulk and McLean (1978) profile gives B = 4.9 mG (29% below our value), while the Pätzold et al. (1987) profile gives B = 2.7 mG (61% below our value). Clearly our profile is closer to the Dulk and McLean (1978) profile than to the Pätzold et al. (1987) profile. The deviation of the profile in eq.(6) from those in eq. (7) and (8) is much smaller at shorter distances: at the first measurement distance, the deviations are much smaller, but in the opposite direction.

Magnetic field estimates from Faraday rotation measurements of the solar corona using the Very Large Array (VLA) at 5 and 6.2 Rs are consistent with our estimates and thus provide additional support for our technique. Spangler (2005) reported B ~ 39 mG at r = 6.2 Rs using observations made in 2003. This height overlaps with our range of measurements: if we use $B(r) = 0.409r^{-1.30}$ (see Fig. 3), we get B = 38 mG, which is nearly identical to the Spangler (2005) value. From another set of measurements made in 2005, Ingleby et al. (2007) reported B in the range 46 – 52 mG (r = 5Rs) and 30 – 34 mG (r = 6.2 Rs). Our curve gives 50 mG (r = 5 Rs) and 38 mG (r = 6.2 Rs), quite consistent with the Ingleby et al. (2007) values. Other curves in Fig. 3 give similar values, differing only by a few mG.

Bemporad and Mancuso (2010) combined white-light and EUV data to B by applying the Rankine–Hugoniot relation to a shock that showed radio, EUV, and white light signatures. They obtained B ~19 mG at r = 4.3 Rs. This is smaller by ~69% compared to the value (61 mG) given by our radial profile at this distance. These authors attribute the smaller value to the high-latitude corona where they made the measurement. As pointed out by Dulk and McLean (1978), the magnetic field and density in the corona can vary from one active region to another by an order of magnitude. We have already identified a large set of CME events that do show white-light shock structure (Kim et al., 2011, under preparation). These events are being analyzed to understand the extent to which the coronal magnetic field may vary.

## 4. Discussion and Conclusions

The primary finding of this paper is that the CME shock structure identified in coronagraphic observations can be used to estimate the magnetic field strength and its variation with heliocentric distance. The density and magnetic field values determined here can constrain the coronal plasma beta, which is important in understanding the coronal dynamics at large distances from the Sun. We combined data from STEREO and SOHO observations for the same CME because it was a limb event for both the spacecraft. It is remarkable that the results are consistent given that the SOHO and STEREO coronagraphs have different sensitivities, and view the CME at different angles (the separation between SOHO and STEREO was ~24$^o$ at the time of the observations). It is generally difficult to measure the magnetic field in this part of the corona, so the technique presented here represents a significant improvement of the situation. This technique also extends the magnetic field profile to larger distances (23 Rs compared to 10 Rs by Dulk and McLean, 1978 and 15 Rs by Pätzold et al., 1987). It must be possible to extend the measurement to greater heliocentric distances if one can distinguish the shock and CME structures in the heliospheric imager (HI) data from STEREO. One has to systematically examine the HI data for shock-driving events to identify shock structures. Direct measurements of the magnetic field is expected in the future when the magnetometers on board NASA's Solar Probe Plus mission probes the corona in the spatial domain considered here.

The low Mach numbers are consistent with the fact that the shock became radio quiet (the radio type II burst ended, but the shock continued to be observed in white light) at r ~3.7 Rs (Gopalswamy et al., 2009). The standoff distance was measured at the shock nose, where the magnetic field of the ambient medium is expected to be substantially radial and hence the shock quasi-parallel. The decline in Alfven speed as a function of r is also slower than what is expected from empirical models (Gopalswamy et al., 2001; Mann et al., 2003), which give an Alfven speed gradient of ~25 km/s per Rs in the coronal region of interest in this paper. The derived Alfven speeds in Table 1 gives only ~10 km/s per Rs. Note that the model profiles assume both magnetic field and density variation, whereas no such assumption is made here in deriving the Alfven speed profile. However, we do assume the speed profile of the slow solar wind in deriving the Alfven speed.

In conclusion, the new technique for measuring the coronal magnetic field in the outer corona and near-sun interplanetary medium provides an independent means, apart from the Faraday rotation technique. The radial profile of the magnetic can be represented by a power law of the form $B(r) = pr^{-q}$. The curve with p =0.409 and q = 1.30 is in close agreement with published profiles from other techniques, and shows that the magnetic field declines from 48 to 8 mG in the distance range 6 – 23 Rs.

We thank P. Mäkelä and H. Xie for verifying the height-time measurements. This work was supported by NASA's LWS TR&T program. We thank the SOHO and STEREO teams for making the data available online. We thank the referee for helpful comments.

REFERENCES

Bemporad, A. & Mancuso, S. 2010, Astrophys. J., 720, 130

Brueckner, G. E., et al. 1995, Sol. Phys. 162, 357

Cho, K.-S., Lee, J., Gary, D. E., Moon, Y.-J., & Park, Y. D. 2007, Astrophys. J., 665, 799

Dulk, G. A. & McLean, D. J. 1978, Solar Phys., 57, 279

Eselevich, M. V. & Eselevich, V. G. 2011, Astronomy Reports, 55, 359

Gary, G. A. 2001, Solar Phys., 203, 71

Gopalswamy, N. 2009, in Climate and Weather of the Sun-Earth System (CAWSES): Selected Papers from the 2007 Kyoto Symposium, Ed. T. Tsuda, R. Fujii, K. Shibata, and M. A. Geller, pp. 77-120

Gopalswamy, N., Lara, A., Kaiser, M.L., & Bougeret, J.-L. 2001, J. Geophys. Res. 106, 25261.

Gopalswamy, N., Thompson, W. T., Davila, J., Kaiser, M. L., Yashiro, S., Mäkelä, P., Michalek, Bougeret, J.-L., & Howard, R. A. 2009, Solar Phys., 259, 227

Howard, R.A., Moses, J.D., Vourlidas, A., Newmark, J.S., Socker, D.G., Plunkett, S.P., Korendyke, C.M., et al. 2008, Space Sci. Rev. 136, 67

Ingleby, L. D., Spangler, S. R., & Whiting, C. A. 2007, Astrophys. J., 668, 520

Kaiser, M.L., Kucera, T.A., Davila, J.M., St. Cyr, O.C., Guhathakurta, M., Christian, E. 2008, Space Sci. Rev. 136, 5

Lee, J. 2007, Space Sci. Rev. 133, 73

Lin, H., Penn, M. J., and Tomczyk, S. 2000, Astrophys. J., 541, L83

Mann, G., Klassen, A., Aurass, H., & Classen, H.-T. 2003, Astron. Astrophys. 400, 329.

Ontiveros, V. & Vourlidas, A. 2009, Astrophys. J., 693, 267

Pätzold, M., Bird, M. K., Volland, H., Levy, G. S., Siedel, B. L., and Stelzried, C. T. 1987, Solar Phys., 109, 91.

Russell, R. T. & Mulligan, T. 2002, Planetary and Space Science 50, 527

Saito, K., Poland, A.I., & Munro, R.H. 1977, Solar Phys. 55, 121

Sheeley, N. R., Jr., et al. 1997, Astrophys. J., 484, 472

Sheeley, N. R., Jr., Hakala, W. N., & Wang, Y.-M. 2000, J. Geophys. Res., 105, 5081

Spangler, S. R. 2005, Space Sci. Rev., 121, 189

Thernisien, A. F. & Howard, R. A. 2006, Astrophys. J., 642, 523

Vourlidas, A., Wu, S. T., Wang, A. H., Subramanian, P., & Howard, R. A. 2003, Astrophys. J., 598, 1392

Wiegelmann, T. 2008, J. Geophys. Res., 113, A03S02